\documentclass[prl,twocolumn,showpacs]{revtex4-1}

\usepackage{amsmath}
\usepackage{bm}
\usepackage{graphicx}
\usepackage{epstopdf}
\usepackage{hyperref}

\newcommand{\OCDM}{\Omega_{\rm c}}

\newcommand{\MPl}{M_{\rm Pl}}
\newcommand{\LQCD}{\Lambda_{\rm QCD}}

\usepackage{color}

\begin{document}

\title{Axion cold dark matter in view of BICEP2 results}

\author{Luca Visinelli}
\email[Electronic address: ]{visinelli@utah.edu}
\author{Paolo Gondolo}
\email[Electronic address: ]{paolo.gondolo@utah.edu}
\affiliation{Department of Physics, University of Utah, 115 S 1400 E $\#$201, Salt Lake City, UT 84102, USA.}
\date{\today}

\begin{abstract}
The properties of axions that constitute 100\% of cold dark matter (CDM) depend on the tensor-to-scalar ratio $r$ at the end of inflation. If $r=0.20^{+0.07}_{-0.05}$ as reported by the BICEP2 collaboration, then ``half'' of the CDM axion parameter space is ruled out. Namely, the Peccei-Quinn symmetry must be broken after the end of inflation, and axions do not generate non-adiabatic primordial fluctuations.  The cosmic axion density is then independent of the tensor-to-scalar ratio $r$, and the axion mass is expected to be in a narrow range that however depends on the cosmological model before primordial nucleosynthesis. In the standard $\Lambda$CDM cosmology, the CDM axion mass range is  $m_a = \left(71 \pm 2\right) \mu{\rm eV} \, (\alpha^{\rm dec}+1)^{6/7}$, where $\alpha^{\rm dec}$ is the fractional contribution to the cosmic axion density from decays of axionic strings and walls.
\end{abstract}


\maketitle


Precision cosmological measurements \cite{komatsu,planck} have established the relative abundance of dark and baryonic matter in our Universe. About 84$\%$ of the content in the Universe is in the form of cold dark matter (CDM), whose composition is yet unknown. One of the most promising hypothetical particles proposed for solving the enigma of the dark matter nature is the axion \cite{weinberg,wilczek}. Axions were first considered in 1977 by Peccei and Quinn (PQ \cite{peccei}) in their proposal to solve the strong-CP problem in quantum chromodynamics (QCD). For this purpose, they introduced a U(1) symmetry that is spontaneously broken below an energy scale $f_a$. Although the original PQ axion with $f_a$ around the electroweak scale was soon excluded, other axion models (`invisible' axions) are still viable \cite{Kim:1979if, shifman, dine, zhitnitskii}. Astrophysical considerations on the cooling time of white dwarfs yield the bound \cite{raffelt1}
$
f_a > 4 \times 10^8\, {\rm GeV},
$ 
valid for KSVZ axions. A similar bound from supernovae applies to other axion models \cite{raffelt1}.

The hypothesis that the axion can be the dark matter particle has been studied in various papers (see e.g. \cite{preskill, abbott, dine1, stecker, wilczek1, lyth, beltran, hertzberg} and the reviews in \cite{fox, sikivie}). In particular, in \cite{visinelli1,visinelli2} we examined the axion parameter space for the important case in which axions account for the totality of the observed CDM. We concluded that in the standard $\Lambda$CDM cosmology, the CDM axion mass $m_a$ can theoretically be either in the wide mass range $\sim 10^{-12}$--$10^{-2}$ eV (if the PQ symmetry breaks before the end of inflation), or in the narrow mass range $(\alpha^{\rm dec}+1)(85\pm3)$ $\mu$eV (if the PQ symmetry breaks after the end of inflation; here $\alpha^{\rm dec}$ is the fractional axion density from decays of axionic topological defects, contentiously argued to be $\sim 0.2$, $\sim 10$, or $\sim 200$ --- see discussion at the end of the next Section).

In this note we remark that the measurement \cite{BICEP2} of a tensor-to-scalar ratio $r=0.20^{+0.07}_{-0.05}$ in the cosmic microwave background excludes the first possibility (PQ symmetry breaks before the end of inflation), and thus restricts the CDM axion mass to a narrow range that begs to be located through improved studies of axionic string decays. 

As in \cite{visinelli1,visinelli2}, we impose throughout the requirement that the axion energy density equals the total cold dark matter density,
\begin{equation} \label{CDM}
\Omega_a h^2 = \OCDM h^2 = 0.1196 \pm 0.0031 \quad \text{at 68\% CL \cite{planck}}.
\end{equation}
Here $\Omega_a$ and $\OCDM$ are the densities of axions and of cold dark matter in units of the critical density $\rho_c = 3H^2_0 \,\MPl^2/8\pi$, where $M_{\rm Pl}=1.221 \times 10^{19}$ GeV is the Planck mass, and $h$ is the Hubble constant $H_0$ in units of 100\,km s$^{-1}$Mpc$^{-1}$.

The phenomenology of axion CDM depends on the value of the Hubble expansion rate $H_I$ at the end of inflation, which can be obtained from the tensor-to-scalar ratio $r$ and other CMB data as follows. The curvature perturbation spectrum $\Delta^2_{\mathcal{R}}(k_0)$ at fixed wave number $k_0 = 0.002\, {\rm Mpc}^{-1}$ has been measured at 68\%CL as \cite{komatsu}
\begin{equation}\label{measure_spectrum}
\Delta^2_{\mathcal{R}}(k_0) = (2.445 \pm 0.096)\times 10^{-9}.
\end{equation}
(Planck \cite{planck} reports a similar $\Delta^2_{\mathcal{R}}(k_0)=A_s=(2.21 \pm 0.1)\times 10^{-9}$ at $k_0=0.05\, {\rm Mpc}^{-1}$.)
Similarly, tensor modes have spectrum
\begin{equation}
\Delta^2_h(k_0) = \frac{2 H^2_I}{\pi^2\, M^2_{\rm Pl}}.
\end{equation}
In terms of the tensor-to-scalar ratio
\begin{align}
r_{k_0} = \frac{\Delta^2_h(k_0)}{\Delta^2_{\mathcal{R}}(k_0)},
\end{align}
we have
\begin{equation}
H^2_I = \frac{\pi^2}{2} M^2_{\rm Pl}\,\Delta^2_{\mathcal{R}}(k_0) \, r_{k_0}.
\end{equation}
Using the BICEP2 result \cite{BICEP2} 
\begin{equation}
r_{0.002} = 0.20^{+0.07}_{-0.05},
\end{equation}
gives
\begin{equation}
H_I = \left(6.00 \pm 0.91\right) \times 10^{14}\,{\rm GeV}.
\label{HI}
\end{equation}
We remark that there seems to be tension between the BICEP value of $r$ and the WMAP upper limit $r<0.11$ at 95\%, but as discussed in the BICEP2 paper, this tension is model-dependent and can be alleviated in some models (e.g., with a running spectral index).

\section{Axion CDM}

Axions are the quanta of the axion field $a(x)$ \cite{weinberg, wilczek1}, which is the phase of the PQ complex scalar field after the spontaneous breaking of the PQ symmetry gives it an absolute value $f_a$. Since the U(1) vacuum is topologically a circle, topological defects in the form of axionic strings form at the time of the PQ symmetry breaking. Later, at the time of the QCD phase transition ($T\sim 10^2$ MeV), QCD instanton effects generate an axion potential
\begin{equation}
V(\theta) = m_a^2(T)f_a^2(1-\cos\theta),
\end{equation}
where $\theta(x) = a(x)/f_a$ and $m_a(T)$ is the temperature-dependent axion mass, approximately equal to \cite{gross}
\begin{equation} \label{axion_mass}
m_{a}(T) =
\begin{cases}
m_a, & \text{for~} T\lesssim \LQCD,
\\
b m_a \big(\LQCD/T\big)^4 ,& \text{for~} T\gtrsim \LQCD.
\end{cases}
\end{equation}
Here $b=0.018$~\cite{beltran, fox, hertzberg, visinelli1} and $\LQCD = 200$ MeV \cite{kolb}.
The (zero-temperature) axion mass $m_a$ is \cite{weinberg}
\begin{equation}
\label{eq:axionmass}
m_a = \frac{\sqrt{z}}{1+z}\frac{f_{\pi}m_{\pi}}{f_a/N} = 6.2 \, {\rm \mu eV}\, \bigg( \frac{10^{12}{\rm GeV}}{f_a/N}\bigg),
\end{equation}
where $z \simeq 0.56$ and $m_{\pi}$ and $f_{\pi}$ are the pion mass and decay constant respectively. We choose the color anomaly index $N=1$~\cite{sikivie}.

As the universe expands, two different scenarios occur for cosmic axion production, depending on whether the PQ symmetry breaks before (Scenario B) or after (Scenario A) inflation ends (\cite{hertzberg, kolb, sikivie, fox} and references therein).

In Scenario B, which occurs for 
\begin{align}
f_a > H_I/(2\pi) \simeq 0.95\times10^{14}\,{\rm GeV}
\end{align}
(using the BICEP2 result for $H_I$ in Eq.~(\ref{HI})), axionic topological defects are inflated away and play no role. The axion potential drives coherent field oscillations with a single initial misalignment angle $\theta_i$ over the observable universe. Their energy density appears as cosmic axion energy density.

In Scenario A, which occurs for \begin{align}
f_a < H_I/(2\pi) \simeq 0.95\times10^{14}\,{\rm GeV},
\end{align}
as the universe expands, axionic strings break into axion-radiating closed loops and eventually dissolve into axions. The axion potential drives coherent oscillations with different initial angles $\theta_i$; their energy density must be averaged over a Hubble volume. The cosmic axion energy density contains contributions from string decays and from coherent oscillations (vacuum realignment).

In the vacuum realignment mechanism, the equation of motion for the misalignment angle $\theta = a/f_a$ is
\begin{equation}\label{eq_motion}
\ddot{\theta} + 3H(T)\,\dot{\theta}
+\frac{1}{f_a^2}\frac{\partial V(\theta)}{\partial\, \theta} = 0,
\end{equation}
where a dot indicates a derivative with respect to time, $H(T) = 1.66\sqrt{g_*(T)}\,T^2/\MPl$ is the Hubble rate during the radiation-dominated epoch. For small $\theta$, the potential is approximately harmonic, $V(\theta)\approx\frac{1}{2}m^2_a(T)f^2_a\theta^{2}$, and Eq.~(\ref{eq_motion}) is approximated by
\begin{equation}\label{eq_motion1}
\ddot{\theta} + 3H(T)\,\dot{\theta} + m^2_a(T)\theta = 0.
\end{equation}
When $T \gg \LQCD$, the axion is massless, and Eq.~(\ref{eq_motion1}) is solved by $\dot\theta=0$, $\theta = \theta_i(x)$, where $\theta_i(x)$ is the initial misalignment angle, which generally depends on position. The axion field is frozen at the value $\theta_i$ until a temperature $T_f$ at which
\begin{equation} \label{def_T1}
3H(T_f) = m_a(T_f).
\end{equation}
Using the axion mass in Eq.~(\ref{eq:axionmass}), and assuming a standard radiation-dominated cosmology before primordial nucleosynthesis, we find \cite{visinelli1}
\begin{equation} \label{T_f}
T_f =
\begin{cases}
618\, {\rm MeV}\, \bigg(\frac{10^{12}{\rm GeV}}{f_a}\bigg)^{1/6}, & T \gtrsim \LQCD,\\
68.1 \, {\rm MeV}\,  \bigg(\frac{10^{18}{\rm GeV}}{f_a}\bigg)^{1/2}, & T \lesssim \LQCD.\\
\end{cases}
\end{equation}

The misalignment mechanism contributes a cosmic axion density at temperature $T_f$ equal to \cite{kolb, sikivie, visinelli1,visinelli2}
\begin{equation} \label{number_nonstandard}
n_a(T_f) = \frac{1}{2}\,\chi\,m_a(T_f)\,f_a^2\langle\,\theta_i^2\,f(\theta_i)\rangle.
\end{equation}
Here, $\chi$ is a model-dependent factor that depends on the number of quark flavors $N_f$ that are relativistic at $T_f$ \cite{turner}, while the function $f(\theta_i)$ accounts for anharmonicity in the axion potential, i.e., for a solution to the full axion field Eq.~(\ref{eq_motion}) instead of Eq.~(\ref{eq_motion1}) \cite{turner,lyth1,strobl,bae,kobayashi}. Here, we set $\chi = 1.44$, consistent with $N_f =3$, and we consider the analytic anharmonicity function in Ref.~\cite{visinelli1},
\begin{equation} \label{anharmonicity_function}
f(\theta_i) = \left[\ln\left(\frac{e}{1-\theta_i^2/\pi^2}\right)\right]^{7/6}.
\end{equation}

The axion number density $n_0$ at the present time is found by conservation of the comoving axion number density,
\begin{equation} \label{energydensity}
n_0 =
\frac{m_a(T_f)\,\chi\,s(T_0)}{2s(T_f)}f_a^2\langle\theta_i^2\,f(\theta_i)\rangle,
\end{equation}
where the entropy density with $g_{*S}(T)$ degrees of freedom at temperature $T$ is
\begin{equation} \label{entropy}
s(T) = \frac{2\pi^{2}}{45}\,g_{*S}(T)\,T^3.
\end{equation}
The present cosmic axion mass density $\rho_a = m_a\,n_0$ from vacuum misalignment follows as, taking $g_{*}$ as in \cite{visinelli1},
\begin{equation} \label{standarddensity}
\Omega_a^{\rm mis} h^2 =
\begin{cases}
0.236\,\langle\theta_i^2\,f(\theta_i)\rangle(f_{a,12})^{7/6}, & f_a \lesssim \hat{f}_a,\\
0.0051\,\langle\theta_i^2\,f(\theta_i)\rangle(f_{a,12})^{3/2}, & f_a \gtrsim \hat{f}_a.\\
\end{cases}
\end{equation}
where $\hat{f}_a = 0.991\times 10^{17}{\rm GeV}$ and $f_{a,12}=f_a/10^{12}\,{\rm GeV}$.  

The angle average $\langle\theta_i^2f(\theta_i)\rangle$ assumes different values in Scenario A and Scenario B. In Scenario B, the initial misalignment field $\theta_i$ is uniform over the entire Hubble volume, but there are axion quantum fluctuations of variance $\sigma_\theta^2$ arising from inflation, so
\begin{equation} \label{theta_scenario_II}
\langle\theta_i^2\,f(\theta_i)\rangle = \left( \theta_i^2 + \sigma_{\theta}^2 \right)\, f(\theta_i).
\end{equation}
Since at this stage the axion is practically massless, its quantum fluctuations have the same variance as the inflaton fluctuations \cite{birrell},
\begin{equation} \label{standard_deviation}
\sigma^2_{\theta} = \bigg(\frac{H_I}{2\pi f_a}\bigg)^2.
\end{equation}
Hence in Scenario B, since there is no contribution to the cosmic axion density from decays of axionic topological defects, the total axion energy density is given by
\begin{equation} \label{omegaB}
\Omega_a h^2 =
\begin{cases}
0.236\,\big[\theta_i^2+\left(\frac{H_I}{2\pi f_a}\right)^2\big]f(\theta_i)(f_{a,12})^{7/6}, & f_a \lesssim \hat{f}_a,\\
0.0051\,\big[\theta_i^2+\left(\frac{H_I}{2\pi f_a}\right)^2\big]f(\theta_i)(f_{a,12})^{3/2}, & f_a \gtrsim \hat{f}_a.\\
\end{cases}
\end{equation}

In Scenario A, the variance of the axion field is zero because there are no axion quantum fluctuations from inflation, but $\theta_i$ is not uniform over a Hubble volume, so $\theta_i^2$ is averaged over its possible values as~\cite{visinelli1}
\begin{equation}
\langle\theta_i^2\,f(\theta_i)\rangle = \frac{1}{2\pi}\int_{-\pi}^{\pi}\theta_i^2 \, f(\theta_i) \, d\theta_i = 2.67\,\frac{\pi^2}{3}.
\end{equation}
Hence, from Eq.~(\ref{standarddensity}), since $f_a < \hat{f}_a$ in Scenario A,
\begin{equation}
\Omega_a^{\rm mis} h^2 = 2.07\,\left(f_{a,12}\right)^{7/6} \quad \text{(Scenario A).}
\end{equation}
Extra contributions $\Omega_a^{\rm dec}$ from decays of axionic topological defects are present in Scenario A. Their calculation requires difficult numerical simulations of particle production from axionic strings and walls evolving in the expanding universe. Results have been discrepant and controversial for decades. They can be expressed as ratios $\alpha^{\rm dec} = \Omega_a^{\rm dec}/\Omega_a^{\rm mis}$ of topological-defect decay densities to vacuum realignment densities. For example, Refs.~\cite{harari,hagmann}, Refs.~\cite{hiramatsu}, and Refs.~\cite{davis,battye} find string-to-misalignment ratios of $\sim 0.16$, $\sim 6.9\pm3.5$, $\sim186$, respectively, while Ref.~\cite{hiramatsu} argues for a combined wall-and-string-to-misalignment ratio $\alpha^{\rm dec} \sim 19 \pm 10$ (see \cite{visinelli2,hiramatsu} for further references). Including the contributions from decays of axionic topological defects,
\begin{equation}
\Omega_a h^2 = (\alpha^{\rm dec}+1) \, 2.07\,\left(f_{a,12}\right)^{7/6} \quad \text{(Scenario A).}
\end{equation}

\section{Constraints}

\begin{figure}[t]
\includegraphics[width=0.48\textwidth]{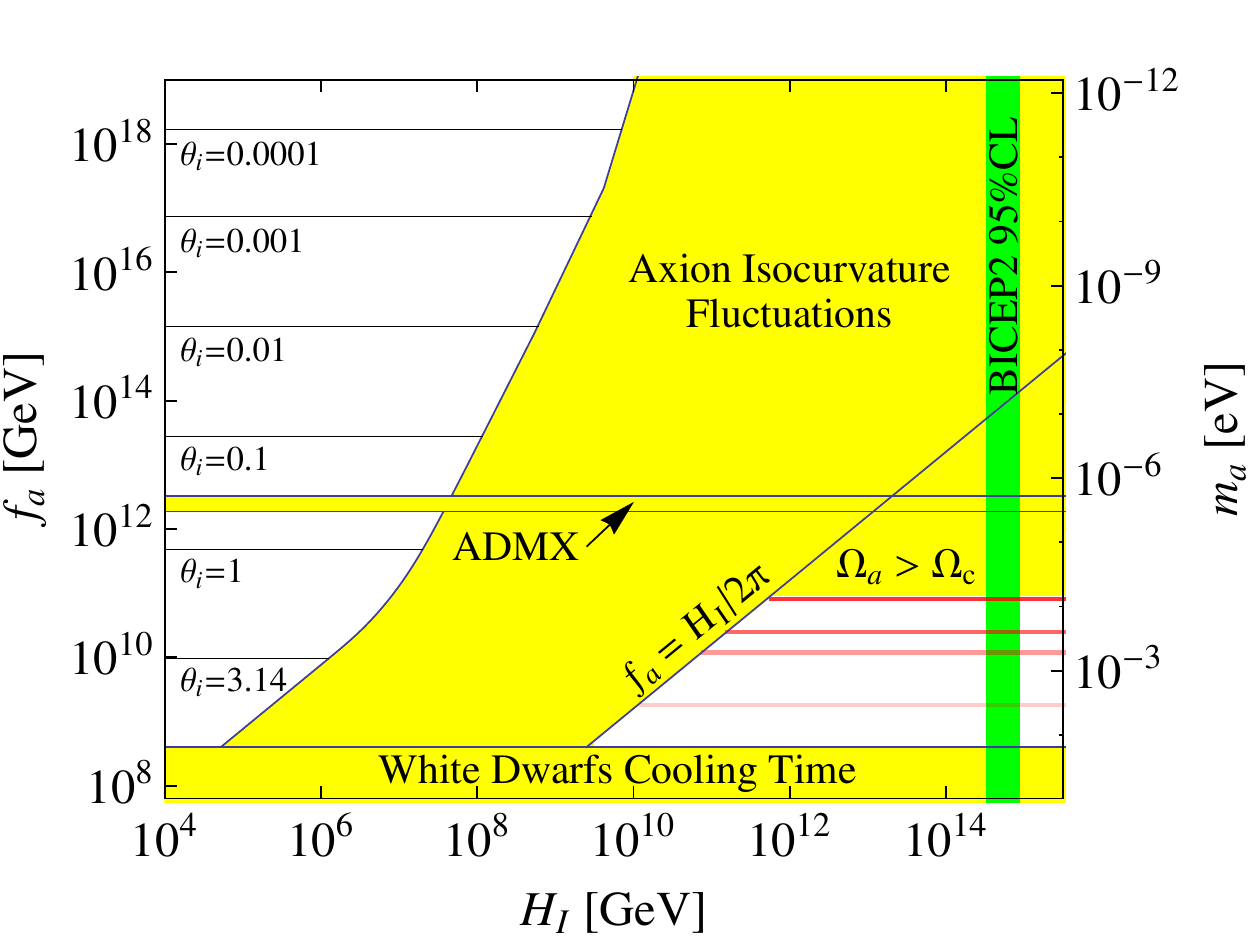}
\caption{CDM axion parameter space. Yellow regions: excluded. Green band: BICEP2 measurement of $r$. Colored horizontal bands: $\Omega_a=\Omega_{\rm c}$ for some models of axion production by decays of axionic topological defects. The BICEP2 measurement excludes Scenario B ($f_a>H_I/2\pi$). The intersection of the colored bands shows the preferred CDM axion masses.}
\end{figure}

Figure 1 shows a summary of the constraints on the CDM axion parameter space $H_I$--$f_a$, showing a complete range for $f_a$ up to the Planck scale. Shaded in yellow are all regions excluded before the BICEP measurement (with the omission of the WMAP upper limit on $r$). Axions could have been 100\% of CDM in the white region on the left (Scenario B) and in one of the narrow colored horizontal bands on the bottom right, which represent the $\Omega_a = \Omega_{\rm c}$ condition for the four examples of axionic string-wall decays mentioned above (Scenario A). The BICEP2 reported measurement of $r$ is indicated by the green vertical band. Clearly the BICEP2 measurement excludes Scenario B.

The main constraint on Scenario B comes from non-adiabatic fluctuations in the axion field, which are constrained by WMAP measurements. The power spectrum of axion perturbations $\Delta_a^2(k) = \langle|\delta \rho_a/\rho_a|^2\rangle$ is given by
\begin{equation} \label{axion_perturbations}
\Delta^2_a(k) = \frac{H^2_I}{\pi^2 \theta^2_i f^2_a}.
\end{equation}
Hence
\begin{equation}
\frac{\Delta^2_a(k_0)}{\Delta^2_{\mathcal{R}}(k_0)} =
\frac{H_I^2}{\pi^2 \Delta^2_{\mathcal{R}}(k_0) \theta_i^2 f^2_a} = \frac{\alpha_0(k_0)}{1-\alpha_0(k_0)},
\end{equation}
where the axion adiabaticity $\alpha_0(k_0)$ is constrained by the WMAP 5-year data~\cite{komatsu} to
\begin{equation} \label{alpha}
\alpha_0 < 0.072\quad\quad \text{(at 95\% CL).}
\end{equation}
Using the value of $\Delta^2_{\mathcal{R}}(k_0)$ in Eq.~(\ref{measure_spectrum}) and the BICEP2 result for $H_I$ in Eq.~(\ref{HI}), this bound can be rephrased as
\begin{equation} \label{adiabaticity}
\theta_i \,f_{a,12} > 2.8 \times 10^{11}.
\end{equation}
Combined with Eq.~(\ref{omegaB}), this leads to the bounds
\begin{gather}
\theta_i < \left(\frac{\OCDM\,h^2}{0.0051}\right)^2\,\left(\frac{10^8{\rm ~GeV}}{2.4\,H_I}\right)^3 = 1.65 \times 10^{-19},
\\
f_a > 9\times 10^{37}{\rm~GeV}.
\end{gather}
The latter is much larger than the Planck scale and therefore Scenario B is excluded.

Scenario A extends in the region $f_a < H_I/2\pi$, which for the BICEP2 value of $H_I$ corresponds to
\begin{equation}
f_a < 0.96 \times 10^{14}\,{\rm ~GeV}, 
\quad
m_a > 0.06\,{\rm~\mu eV}.
\end{equation}
In this scenario, the axion energy density does not depend on the tensor-to-scalar ratio $r$. The preferred PQ scale and mass for CDM axions depend on the contribution $\alpha^{\rm dec}$ from decays of axionic strings and walls. We find them to be
\begin{gather}
f_a = \left(8.7 \pm 0.2\right) \times 10^{10}{\rm ~GeV} \, (\alpha^{\rm dec}+1)^{-6/7},
\\
m_a = \left(71 \pm 2\right) \mu{\rm eV} \, (\alpha^{\rm dec}+1)^{6/7}.
\end{gather}

Since $\Omega_a h^2 \le \Omega_a^{\rm mis} h^2 \le \Omega_{\rm c} h^2$, the numerical coefficients also represent a cosmological upper limit on $f_a$ and lower limit on $m_a$.

P.G. was supported in part by NSF grant PHY-1068111. Within a day of this note, Refs.~\cite{higaki,marsh} appeared on the same topic.

\end{document}